# The study on penetration depth of anisotropic two-band s-wave superconductors by Ginzburg-Landau approach.

P. Tongkhonburi[a,b] and P. Udomsamuthirun[a]

[a]Prasarnmit Physics Research Unit, Department of Physics, Faculty of Science, Srinakharinwirot University, Sukumvit23 Road, Bangkok 10110, Thailand.

[b]Silverlake Technology Co., Ltd. Bangsaotong, Samutprakarn 10570, Thailand.

Email: pongkan_sl@yahoo.com

Abstract

The penetration depth of anisotropic two-band s-wave superconductors have been studied by Ginzburg-Landau approach. The anisotropic spherical Fermi surfaces and order parameters with the spherical ellipse or pancake Fermi shape are used for derive penetration depth equation. The numerical calculation of pancake-pancake and pancake-ellipse Fermi sphere shape are fits well with the experiment data of $MgB_2$ and $CaAlSi$ superconductor, consequently. Our model should be the way to increase understanding in the microscopic mechanism of occurrence of the penetration depth in superconductivity state of these material.

## 1.Introduction

The discovery of two-band superconductor was attractive attention for theoretical frame work in microscopic level. The two-band superconductor has shown the diverge properties from one-band superconductor. There are many kinds of two-band superconductor found such as Magnesium Diboride $MgB_2$, Strontium Ruthenate $Sr_2RuO_4$, nonmagnetic borocarbides $LuNi_2B_2C$, $YNi_2B_2C$ and Ferropnictides. The $MgB_2$ superconductor is the most favorite two-band superconductor that shown the s-wave superconductor with Fermi surface related to σ-bands and π-bands [1,2]. MgB$_2$ superconductor is the type II superconductor with estimated valued at zero temperature of coherence length ξ=5.2 nm, London penetration depth λ=125-140 nm [3,4]. Askerzade and Gencer [5] studied the London penetration depth of MgB$_2$ superconductor by using the two-band Ginzburg-Landau theory and apply to determine the temperature dependence of the London penetration depth. Their results shown that the two-band GL theory can describe well with the experimental data of London penetration depth for MgB$_2$ superconductor. Konstantin and Grigorishin [6] applied the GL theory of the isotropic multiband superconductor for analytic calculation of the London penetration depth, a coherence length and GL parameter. They also showed how to reduce the free energy functional of multiband superconductor to effective GL free energy of single-band superconductor. The Sr$_2$RuO$_4$ superconductivity was discovery with low critical temperature, T$_c$ = 1.5 K [7] and p-wave superconductor [8]. However, Maki et al. [9] proposed the gap symmetries in Sr$_2$RuO$_4$ that be the f-wave superconductor. The penetration depth was also studied in semiclassical model [10]. This approach provided a method for calculating three spatial component of penetration depth of singlet pairing state. Prozorov et al. [11] studied the in-plane and out-of-plane London penetration depth of CaAlSi superconductor by semiclassical approach. The CaAlSi has shown the s-wave superconductor with its properties deviate from a single isotropic gap. The single crystal of CaAlSi were synthesis with 2 values of the critical temperature at 6.2 K and 7.3 K depending on stacking sequences [12]. They assume the superconducting gap is isotropic in ab-plane and anisotropic for out-of-plane and the superfluid density of this superconductor was calculated in BCS scenario and comparing to the experimental data.

In this paper, we are interested in penetration depth of the superconductivity on MgB$_2$ and CaAlSi that are the two-band and anisotropic band s-wave superconductor. The two-band Ginzburg-Landau approach with difference anisotropic Fermi sphere shapes were used for calculating penetration depth to comparing to the experimental data.

**2. Calculation**

We used the Ginzburg-Landau approach for our calculation. Here, we make the assumption that both superconductor MgB$_2$ and CaAlSi having the two-band superconductor properties that they can use the two-band Ginzburg-Landau method. For two-band Ginzburg-Landau theory, there are two order parameters in the Ginzburg-Landau functional that free energy be written as

$$F_{sc}[\psi_1,\psi_2] = \int d^3 r f_{sc} = \int d^3 r \left( F_1 + F_2 + F_{12} + \int H_s dB \right) \quad (1)$$

with
$$F_{i(i=1,2)} = \frac{1}{2m_i}\left|(-i\hbar\bar{\nabla} - 2e\vec{A})\psi_i\right|^2 + \alpha_i|\psi_i|^2 + \frac{1}{2}\beta_i|\psi_i|^4 \quad (2)$$

$$F_{12} = \varepsilon(\psi_1^*\psi_2 + c.c.) + \varepsilon_1\left\{(i\hbar\bar{\nabla} - 2e\vec{A})\psi_1^*(-i\hbar\bar{\nabla} - 2e\vec{A})\psi_2 + c.c.\right\} \quad (3)$$

Here, $F_i$ is the free energy of the separate bands. $F_{12}$ is the interaction term between bands. $m_i$ denotes the effective mass of carriers. $\psi_i$ is the order parameter and $|\psi_i|^2$ is proportional to the density of carriers, the coefficient $\alpha_i$ depends on the temperature, while coefficient $\beta_i$ is independent of temperature.

According to the Gorkov's microscropic derivation of Ginzburg-Landau equation, the gap function and order parameter have a relation as $\psi(r) \equiv \sqrt{\frac{7\xi(3)n}{8(\pi T_c)}}\Delta(r)$ where $n$ is the total electron density. The correspondence is valid in the temperature near critical temperature and gap is sufficient small [13]. Then, we make the assumption that $\Delta \approx \psi$ for our consideration. In Ginzburg-Landau approach, the order parameter can include the anisotropic function by setting

$$\Delta \approx \Delta(T) f(k) \quad (4)$$

And anisotropic order parameter is

$$\psi \approx \psi(T) f(k) \quad (5)$$

From Eqs. (1-5), the current density equation can be calculated by taking the derivative with respect to vector-potential as

$$\frac{\partial F_{SC}}{\partial \vec{A}} = \vec{J} = \frac{2}{2m_1}\left|-i\hbar\nabla - 2e\vec{A})\psi_1\right|(-2e\psi_1) + \frac{2}{2m_2}\left|-i\hbar\nabla - 2e\vec{A})\psi_2\right|(-2e\psi_2)$$
$$+ \varepsilon_1\langle f_1(\hat{k})f_2(\hat{k})\rangle\left( (i\hbar\nabla - 2e\vec{A})(-2e\psi_2\psi_1^* - 2e\psi_1^*\psi_2) + (-i\hbar\bar{\nabla} - 2e\vec{A})\psi_1(-2e\psi_2^*) \right) \quad (6)$$

According to Maxwell's equations, we can apply curl operation on the current density equation then

$$\nabla \times \vec{J} = \frac{1}{\mu_0}\nabla \times \nabla \times \vec{B} = [\frac{4e^2|\psi_1|^2}{m_1}\langle f_1^2(\hat{k})\rangle + \frac{4e^2|\psi_2|^2}{m_2}\langle f_2^2(\hat{k})\rangle + \varepsilon_1\langle f_1(\hat{k})f_2(\hat{k})\rangle(8e^2\psi_2\psi_1^* + 8e^2\psi_1\psi_2^*)]\vec{B} \qquad (7).$$

The magnetic field was applied only in z-direction which $\vec{B}_Z = (0,0,\vec{B}(x))$ then $\nabla \times \nabla \times \vec{B} = \frac{d^2\vec{B}(x)}{dx^2}$, and substituting into Eq. (7), we found that

$$\frac{1}{\mu_0}\frac{d^2\vec{B}(x)}{dx^2} - [\frac{4e^2|\psi_1|^2}{m_1}\langle f_1^2(\hat{k})\rangle + \frac{4e^2|\psi_2|^2}{m_2}\langle f_2^2(\hat{k})\rangle + \varepsilon_1\langle f_1(\hat{k})f_2(\hat{k})\rangle 8e^2(\psi_2\psi_1^* + \psi_1\psi_2^*)]\vec{B} = 0 \qquad (8).$$

Rearrange equation into standard form, $\frac{d^2\vec{B}(x)}{dx^2} - \frac{1}{\lambda^2}\vec{B}(x) = 0$, we can get the penetration depth as

$$\frac{1}{\lambda^2} = 16e^2\mu_0\left[\frac{|\psi_1|^2\langle f_1^2(\hat{k})\rangle}{m_1} + \frac{|\psi_2|^2\langle f_2^2(\hat{k})\rangle}{m_2} + \varepsilon_1\langle f_1(\hat{k})f_2(\hat{k})\rangle(\psi_2\psi_1^* + \psi_1\psi_2^*)\right] \qquad (9)$$

Our penetration depth equation, Eq. (9), is agreed with the anisotropic magnetic London penetration depht of magnetic two-band superconductor of Changjan and Udomsamuthirun [14] and was agreed with Askerzade's calculation [15] that the Ginzburg-Landau for two-band s-wave was used for comparison experimental data for the superconducting magnesium diboride, MgB$_2$, and non-magnetic borocarbides LuNi$_2$B$_2$C, YNi$_2$B$_2$C.

Let $\eta = \frac{8e^2\mu_0}{m}$ and we set that both bands having the same mass of the carrier, $2m = m_1 = m_2$ and $\varepsilon_1 = \varepsilon_1'/2m$ Then Eq. (9) was into the form to do a numerical calculation, we get

$$\frac{1}{\lambda^2} = \eta'\left[|\psi_1|^2\langle f_1^2(\hat{k})\rangle + |\psi_2|^2\langle f_2^2(\hat{k})\rangle + \varepsilon_1'\langle f_1(\hat{k})f_2(\hat{k})\rangle(\psi_2\psi_1^* + \psi_1\psi_2^*)\right] \qquad (10).$$

Here $\eta' = \zeta\eta$ and $\zeta$ is the scaling parameter of $\eta$ and $\lambda$. The numerical calculation of anisotropic gap function can be applied by using the anisotropic model of ellipse model of Haas and Maki [16]: $\Delta_k = \Delta(0)\frac{1+az^2}{1+a}$, and pancake model of Posazhennikova, Dahm and Maki [17]: $\Delta_k = \frac{\Delta(0)}{\sqrt{1+az^2}}$, where $a$ is the anisotropic parameter and $z = \cos\theta$, $\theta$ is the polar angle with respect to the c-axis. The ellipse and pancake model were used to describe the anisotropic s-wave superconductor in MgB$_2$. In Haas and Maki model [16], they assumed ellipsoidal Fermi surface in weak-coupling limit and describe the energy gap, density of state, specific heat, upper critical field of MgB$_2$ superconductor in one-band s-wave model. And the Posazhennikova,

Dahm and Maki [17] model, they also used the model of anisotropic with the pancake sharp to describe gap-to-Tc ratio, density of state, specific heat, conductivity, upper critical field of MgB2 superconductor in one-band s-wave model. Udomsamuthirun, Peamsuwan and Kumvongsa [18] used the ellipse and pancake model to investigate the specific heat of anisotropic two-band superconductor and applied to the experimental data of MgB2, Lu2Fe3Si5 and Nb3Sn superconductor.

According to Eqs.(4-5), we can get the anisotropic function of the ellipse and pancake model as $f_e(k) = \frac{1+az^2}{1+a}$ and $f_p(k) = \frac{1}{\sqrt{1+az^2}}$, respectively. Then, the spherical Fermi surfaces was affected from anisotropic structure to be anisotropic spherical Fermi surfaces by $f_e(k)$ and $f_p(k)$. Within the penetration depth on Eq.(9), there were $<f^2{}_1(k)>$, $<f^2{}_2(k)>$, $<f_1(k)f_2(k)>$ that implicate the anisotropic function. For generalize, we will apply both types of Fermi surface to both bands of consideration. The set of anisotropic function in two-band superconductors should be in 4 types as following, 1. pancake-pancake $\Delta_1 f_p(k), \Delta_2 f_p(k)$; 2. pancake-ellipse, $\Delta_1 f_p(k), \Delta_2 f_e(k)$; 3. ellipse-pancake $\Delta_1 f_e(k), \Delta_2 f_p(k)$ and 4. ellipse-ellipse $\Delta_1 f_e(k), \Delta_2 f_e(k)$ here $\Delta_1$ and $\Delta_2$ were the zero temperature energy band gap of first band and second band, respectively.

## 3. Results and discussions

The penetration depth of MgB2 superconductor was determined by Jin et al. [19] that the temperature dependence of the magnetic-field penetration depth λ was measured by microwave resonator measurement. The anisotropic two-band gap was found. There were two samples used for measure in this experiment. The first sample yield $T_c = 39K$, $\Delta(0) = 3.8 meV$, $\frac{\Delta(0)}{T_c} = 1.13$, $\lambda(0) = 102 nm$ and the second sample was $T_c = 36K$, $\Delta(0) = 3.2 meV$, $\frac{\Delta(0)}{T_c} = 1.03$, $\lambda(0) = 107 nm$ [19]. Manzano et al. [20] measured the temperature dependence of London penetration depth of MgB2 superconductor with radio frequency technique. They found s-wave pairing with $\frac{2\Delta_s(0)}{T_c} = 1.5$ and $\frac{2\Delta_p(0)}{T_c} = 4.0$. The small gap and large gap are 30 K and 89 K in the polycrystalline data, respectively. Buzea and Yamashita [3] reviewed the properties of MgB2 and two band properties were proposed. The gap-to-$T_c$ ratio with $\frac{2\Delta_s(0)}{T_c} = 1.3$ and $\frac{2\Delta_p(0)}{T_c} = 4.0$ and the small gap and large gap are 2.2 meV and 6.7 meV were shown. To compare our numerical calculation Eq.(10) with the experiment results of λ of MgB2 superconductor [19,20]. The average value of small gap and large gap from the data of Buzea and Yamashita [3] are used to set up our order parameters. Then, we set $|\psi_1| = 25$, $|\psi_2| = 78$, $\eta' = 10000$, $\varepsilon'_1 = 1$, $T_c = 36K$ for calculation. Our parameters are set by considering from the experimental data of MgB2 superconductor with the first gap is a small gap, the second gap is the large gap. There are the scaling factor as $\eta' = 10000$ and a little on the interband tunneling coefficient $\varepsilon'_1 = 1$. After substituting all of parameters into Eq. (10), we can get the relation of London penetration depth versus temperature as Figure 1.

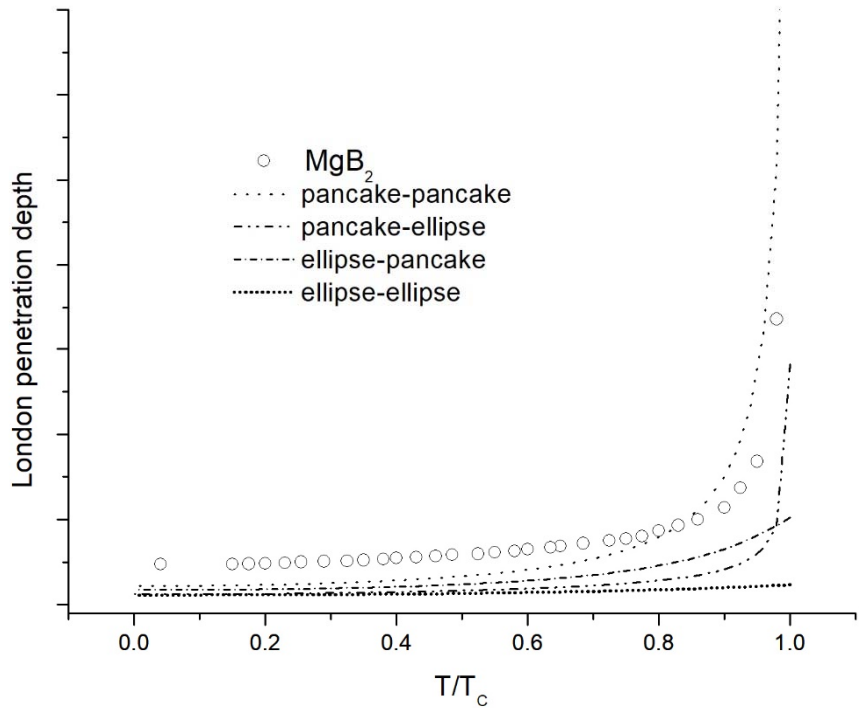

Figure 1. The London penetration depth of our model were fitted to experiment data of MgB2 superconductor.

The λ(0) values of this experiment was in the range of corresponding value from other measurement [21-23]. As T approaches to $T_c$, the penetration depth goes to infinity and the two-band anisotropic with pancake-pancake model can fit well with the experimental data. There is little disagreement between the experimental data and pancake-pancake model for temperature away from the critical temperature. However, our result is ruling out the explanation of Jin et al. [19] that the ellipsoidal Fermi surface had assumed to be the main effect on the penetration depth experiment. Golubov et al. [24] also studied the anisotropy of the magnetic field penetration depth of MgB$_2$ superconductor. They found that there would be kink in the temperature dependence of penetration depth found if there has not the interband scattering. The low and high temperature dependence is determined by smaller gap and larger gap respectively but the smooth line of temperature dependence was found with influence of interband scattering. In our calculation, the parameter of interband effect is found to be very effective parameters to give the results fitted well with the experimental data.

The CaAlSi superconductor is the another example of two-band s-wave superconductor [11]. It has critical temperature about 6-8 K [25] with AlB$_2$ type compound. The $T_c = 6.7K$ with $\frac{2\Delta(0)}{T_c} = 3.53$ and $\frac{2\Delta(0)}{T_c} = 4.22$ and the $T_c = 7.3K$ with $\frac{2\Delta(0)}{T_c} = 3.53$ and $\frac{2\Delta(0)}{T_c} = 3.66$ were found. The energy band with $\frac{2\Delta(0)}{T_c} = 3.53$ is obey with the weak- coupling BCS superconductor. Lupi et al.,[26] report the optical study of CaAlSi superconductor which the two band gap were shown. The resulting value, $2\Delta(0)$, for small and large gap at zero temperature are 8 cm$^{-1}$ and 28 cm$^{-1}$ which $\frac{\Delta_L}{\Delta_s} = 3.5$. The large gap with $\frac{2\Delta(0)}{T_c} = 4.22$ and the small gap with the ratio $\frac{\Delta_L}{\Delta_s} = 3.5$ are used to set up for our calculation . Then, we used the Eq.(10) and set $|\psi_1| = 14$, $|\psi_2| = 4$, $\eta' = 20000$, $\varepsilon'_1 = 1.5$, $T_c = 6.7K$ to be the parameters of calculations. After calculating the anisotropic parameters and substitution into penetration depth, we can compare the all type of anisotropic functions with the experimental of CaAlSi superconductor as shown in Figure 2.

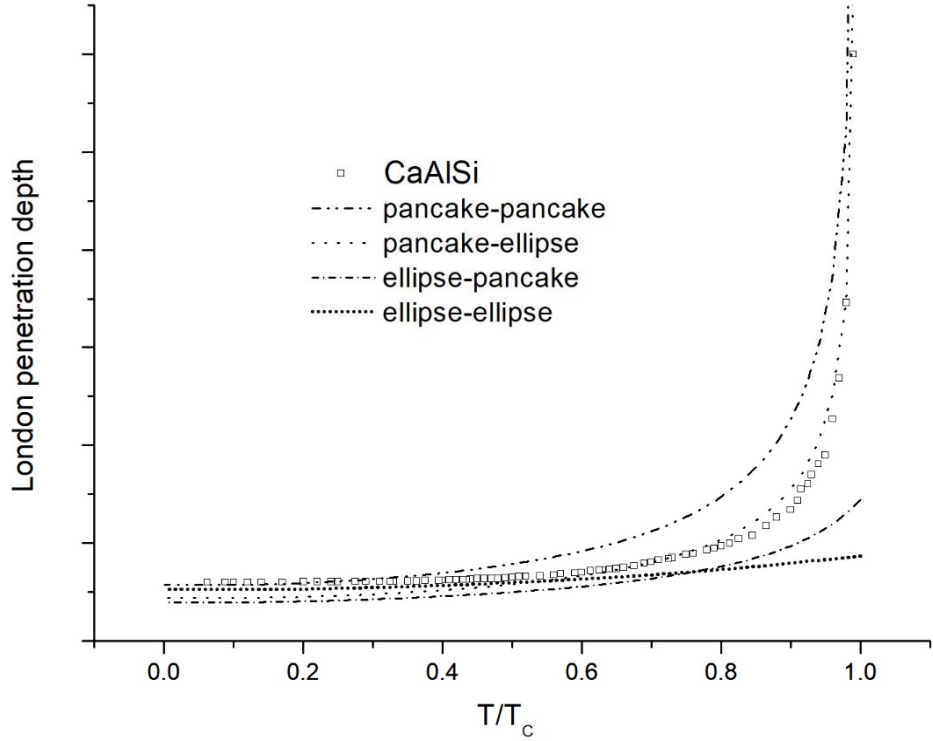

Figure 2. The London penetration depth of our model were fitted to experiment data of CaAlSi superconductor.

We use the data of superfluid density and penetration depth from Prozorov et al.[11] that there were both in ab- and c-direction. Here, the data only in ab-direction is used for comparision to our calculation. The pancake-ellise Fermi surface can fit well with the experimental data. There was the smooth line that the interband scattering was found in our calculation. However, the near critical temperature is shown some difference between experimental data and our calculated resulted.

The our model of calcultions are assumed that the penetration depth of two-band s-wave superconductor is dependence on the shape of Fermi sphere in each band that agree with Chandrasekhar and Einzel [10]. The semiclassical approach is used by Chandrasekhar and Einzel [10] to study the London penetration depth and supercurrent density and they shown the relating the electronic properties to the band structure and Fermi surface. Recently, Chanikul and Udomsamuthirun [27] study the penetration depth and superfluid density of anisotropic one-band s-wave superconductor by semiclassical method and found that the shape of Fermi sphere having a relationship to these properties.

4. Conclusion

The two type of two-band s-wave superconductors, $MgB_2$ and CaAlSi, are investigated in our study. The two-band Ginzburg-Landau approach with difference anisotropic Fermi sphere shapes were used for calculating penetration depth to comparing to the experimental data. The calculation finding represents that the anisotropic of Fermi sphere for pancake-pancake shape can fits well with the experiment data of $MgB_2$ superconductor. And pancake-ellipse Fermi sphere shape is fit to the experiment data of CaAlSi superconductor well. Our model can lead to a way to point out the type of Fermi surface of material that should be the way to increase understanding in the microscopic mechanism of occur some properties in superconductivity state.

Ackknowledgements

Our thanks are to Prasarnmit Physics Research Unit for impeccable and warming support.